# Original mechanism of transformation from soft metallic ($sp^2/sp^3$) $C_{12}$ to ultra-dense and ultra-hard ($sp^3$) semi-conducting $C_{12}$: Crystal chemistry and DFT characterizations.

Samir F. Matar

Lebanese German University (LGU), Sahel Alma, P. O. Box 206 Jounieh, Lebanon.

https://orcid.org/0000-0001-5419-358X. Email: s.matar@lgu.edu.lb

*Dedication*: This work is dedicated to Dr Jean Galy, internationally renowned crystallographer, former Emeritus CNRS 1st Class Research Director at the University of Bordeaux-LCTS (France).

**Abstract.**

*An original mechanism is proposed for a pressure-induced transformation of orthorhombic $C_{12}$ from ground state normal pressure (NP) $sp^2/sp^3$ allotrope to ultra-dense and ultra-hard HP $sp^3$ form. Upon volume decrease, the trigonal C=C parallel segments characterizing glitter-like **tfi** topology of NP $C_{12}$ change to crossing C-C segments with the loss of $sp^2$ character accompanied by a large densification with $\rho$=3.64 g/cm³, larger than diamond, defining a novel orthorhombic HP $C_{12}$ with $4^4T39$ topology. The crystal chemistry engineering backed with quantum density functional theory DFT-based calculations let determine the ground state structures and energy derived physical properties. Furthering on that, the E(V) equations of states (EOS) let define the equilibrium NP($E_0,V_0$) allotrope at lower energy and higher volume versus HP($E_0, V_0$) allotrope at higher energy and smaller volume. A potential pressure induced transformation NP→HP was estimated at ~100 GPa, reachable with a diamond anvil cell DAC. Both allotropes were found cohesive and mechanically stable with low and large Vickers hardness magnitudes: $H_V$(**tfi** $C_{12}$) =24 GPa and $H_V(4^4T39\ C_{12})$ =90 GPa; the latter being close to diamond hardness ($H_V$ ~95 GPa). Besides, both allotropes were found dynamically stable with positive phonon frequencies and a spectroscopic signature of C=C high frequency bands in **tfi** $C_{12}$. The electronic band structures show metallic behavior for NP **tfi** $C_{12}$ and a small band gap for HP $4^4T39\ C_{12}$ letting assign semiconducting properties. The work is meant to open further the scope of $C(sp^2)\rightarrow C(sp^3)$ transformation mechanisms that are fundamental in solid state physics and chemistry.*

**Introduction**

The field of carbon research has a particular position among scientists, with a focus on allotropes related to diamond's physical properties, especially the mechanical and the electronic ones found in artificial allotropes thanks to machined learned programs such as CALYPSO based on evolutionary crystallography [1]. Nevertheless, approaches based on crystal structure engineering rationale also help finding original allotropes, especially with propositions for the transformation from 2D (layered, trigonal) to 3D (three-dimensional, tetrahedral) through puckering and insertion mechanisms [2].

A library of such original -mainly- artificial carbon structures was conceived to store the devised structures, namely SACADA database [3,4]. Therein, C allotropes are described in different topologies identified using TopCryst program [5]. Diamond is known as the hardest material with Vickers hardness amounting to 95 GPa. It is characterized by a high density $\rho \sim 3.55$ g/cm$^3$ arising from the covalent character of the C-C short connections within C(sp$^3$) like tetrahedral carbon. Such hybridization with its inherent chemical properties of electron localization results in a large electronic band gap of ~5 eV. A change to semi-conductive / metallic behaviors useful for applications in electronics, can be achieved by introducing C(sp$^2$) beside C(sp$^3$), leading to electron delocalization. Back in 1994, Bucknum *et al.* [6] proposed hypothetic -with extended Hückel calculations- tetragonal C$_6$ allotrope called "*glitter*", i.e. shining and electron-conductive. The crystal system was built from 1,4-cyclohexadienoid units. The three-dimensional (3D) network is then a crystal system possessing trigonal (sp$^2$) and tetrahedral (sp$^3$) carbon illustrated with brown and white spheres respectively in Fig. 1a. The tetrahedral representation (Fig. 1b) shows further the tetrahedra separated by trigonal C-C (grown spheres). The corresponding C=C -double bond character is illustrated in Fig. 1c with the charge projection yellow volumes calculated presently for the purpose, using the elaborate theory detailed in next section. The density of 3.12 g/cm$^3$ places C$_6$ at an intermediate position between graphite and diamond. Topology wise, glitter C$_6$ was called **tfi** and classified in SACADA database under No. 95 [4]. More recently, another arrangement of cyclohexadiene-type units was identified by us based on diamond-like template structure with an original tetragonal C$_6$ that we called "*neoglitter*" characterized by ultra-hardness with metallic behavior and identified with original **tfa** topology [7].

In this work, starting from crystal structure engineering within **tfi**-like orthorhombic C$_{12}$ characterized by similar stacking of tetrahedra as tetragonal C$_6$ described above, an original ultra-hard orthorhombic allotrope C$_{12}$ was identified through an original mechanism of densification upon volume decrease thanks to the change of trigonal C=C (= meaning double chemical bond $sp^2$-like) parallel segments characterizing glitter-like **tfi** topology of normal pressure NP $C_{12}$ to crossing C-C with complete loss of $sp^2$ character in the high-pressure HP form. The topology of the novel orthorhombic $C_{12}$: $4^4\mathbf{T}39$ characterizing a body center monoclinic $C_{24}$ allotrope [5]. A full study of the physical properties of both allotropes is provided herein.

After this Introduction, the paper is organized as follows: the Theoretical framework and the Computational methodology are given in Section 1; the Crystal Structure characteristics are presented in Section 2; the Mechanical properties from the elastic constants are addressed in Section 3; The Dynamic properties from the Phonons are detailed in Section 4; the pertaining Temperature dependence of the heat capacity in comparison with Diamond experimental data is given in Section 5; Section 6 presents the Electronic band structures. The paper is ended with a Conclusion.

## 1- Theoretical framework and Computational methodology

To determine the ground state structures corresponding to energy minima and to derive the mechanical, dynamic properties and the electronic structures, quantum mechanics computations were carried out based on the widely adopted framework of the density functional theory DFT [8,9]. Based on the DFT, the calculations were performed using the Vienna Ab initio Simulation Package (VASP) code [10,11] and the Projector Augmented Wave (PAW) method [11,12] for the atomic potentials. DFT exchange-correlation (XC) effects were considered using the generalized gradient approximation (GGA) [13]. Relaxation of the atoms to the ground state structures was performed with the conjugate gradient algorithm according to Press *et al*. [14]. The Bloechl tetrahedron method [15] with corrections according to the scheme of Methfessel and Paxton [16] was used for geometry optimization and energy calculations. Brillouin-zone (BZ) integrals were approximated by a special **k**-point sampling according to Monkhorst and Pack [17]. Structural parameters were optimized until atomic forces were below 0.02 eV/Å and all stress components < 0.003 eV/Å$^3$. The calculations were converged at an energy cutoff of 400 eV for the plane-wave basis set in terms of the automatic high precision **k**-point integration in the reciprocal space to obtain a final convergence and relaxation to zero strains for the original stoichiometries presented in this work. In the post-processing of the ground state electronic structures, the charge density projections were operated on the lattice sites.

The mechanical stability was obtained from the elastic constants Cij calculations. The treatment of Cij results was operated thanks to the ELATE online program [18]. The outcome provides the bulk (B) and shear (G) modules along different averaging methods; the Voigt method [19] was used here for $B_V$ and $G_V$. The methods of microscopic theory of hardness by Tian et al. [20] and Chen et al. [21] were used to estimate the Vickers hardness ($H_V$) from the bulk and shear modules $B_V$ and $G_V$. The dynamic stabilities were confirmed by calculating the phonons band structures all presenting positive phonon frequencies. The corresponding phonon band structures were obtained from a high resolution of the orthorhombic Brillouin zone according to Togo *et al*. [22]. The electronic band structures were obtained using the all-electron DFT-based ASW method [23] and the GGA XC functional [13]. The VESTA (Visualization for Electronic and Structural Analysis) program [24] was used to visualize the crystal structures.

2- **Crystal structures and differentiating charge densities**

Orthorhombic $C_{12}$ identified in ground state energy structure with orthorhombic *Ama*2 No. 40 space group is shown in Fig. 2a. As discussed above regarding $C_6$, the structure is characteristic of C-C segments separating tetrahedra as exhibited with the polyhedral projection on the right-hand side. Similarly, orthorhombic $C_{12}$ has **tfi** topology. The crystal parameters are given in Table 1. With an angle below 109.47°, the tetrahedron is distorted because the structure comprises beside *C4* tetrahedra, trigonal carbon C=C -cf. Fig. 1c and charge density projections below. There are two atomic positions, C1 for tetrahedral carbon and C2 for trigonal C. The density is slightly below that of **tfi** $C_6$. The ground state energy given at the bottom of Table 1 let obtain a cohesive carbon system with $E_{coh}$/at.= -2.02 eV deducted from subtracting E(C ) in a large box = -6.6 eV/at.

Original $C_{12}$ in orthorhombic *Aae*2 No. 41 space group shown in Fig. 2b does not exhibit the characteristics of C=C as above. Highlighting the differences with respect to **tfi** $C_{12}$, there can be noted a significant decrease of the cell volume leading to a much larger density of 3.62 g/cm$^3$, higher than diamond's. The tetrahedron angle points to a larger value *versus* **tfi** $C_{12}$ but it remains smaller than the ideal tetrahedron angle. The atomic positions show the same values for tetrahedral C1 but a large x coordinate of C2 (8b) from 0.138 to 0.333 can be highlighted. Then the two structures are close and let propose a transition between them as discussed further below. The cohesive carbon $E_{coh}$/at.= -1.20 eV is higher than **tfi** $C_{12}$. Regarding the topology, $4^4$**T**39 topology was found from TopCryst analyzer, found under No.671 in SACADA database and referring to a monoclinic $C_{24}$ with $I2/m$ No. 12 space group.

**Differentiated charge densities**

The crystal structure differences can be further assessed based on the projection of the charge density. The corresponding volumes are then shown around and between the atomic carbon constituents belonging to the different lattice sites. In Figure 3 the left hand side projection (**tfi** $C_{12}$) clearly shows the differentiated tetrahedral projections versus the trigonal C=C with yellow segments along C-C as in $C_6$ (Fig.1). A completely different configuration appears on the right hand side ($4^4$**T**39 $C_{12}$) where C-C segments lose the $sp^2$ like hybridization forming cross-like segments connecting the tetrahedra. The accompanying decrease of the *c* lattice constant (Table 1) is clearly visible in the projection. It needs to be mentioned that operating such transformation using simpler $C_6$ stoichiometry did not lead to a stable configuration, letting affirm the need of a larger cell as **tfi** $C_{12}$.

3- **Energy-volume equations of state of $C_{12}$ allotropes**

From the crystal structure and energy results we are presented with ground state low energy / high volume **tfi** $C_{12}$ that can be labeled as the normal pressure NP allotrope *versus* high energy / small volume 4,4T39 $C_{12}$ that can then be considered as HP candidate. Such hypotheses are better assessed upon calculating the energy-volume E(V) around ground state values followed by their fit with Birch equations of states (3$^{rd}$ order EOS) letting derive the equilibrium $E_0$, $V_0$ values as well as the bulk modulus $B_0$ and its pressure derivative B'. A series of calculations of the total energy as a function of volume were carried out for the two allotropes. The resulting $E(V)$ curves are shown in Figure 4. Their quadratic fit was carried out to the 3$^{rd}$ order Birch EOS [25]:

$$E(V) = E_0(V_0) + (9/8) \cdot V_0 B_0 [([(V_0)/V]^{2/3} - 1]^2 + (9/16) \cdot B_0 \cdot (B' - 4) \cdot V_0 [([(V_0)/V]^{2/3} - 1]^3,$$

where $E_0$, $V_0$, $B_0$, and $B'$ are the equilibrium energy; the volume; the bulk modulus and its pressure derivative, respectively. The equilibrium values are in reasonable agreement with the energy and volume minima in Table 1.

The fit results in the insert illustrate the above observations.

Using the expression EOS $P = B_0/B'_0\{(V_0/V)^{B'_0} -1\}$, the required pressure for a possible $C_{12}$ **tfi** → $4^4T39$ transition was estimated at P=100 GPa. Such large magnitude could be reached using a diamond anvil cell DAC. Note that the DFT calculations are zero entropy. Then, a raise of temperature can activate the sample and would enable decreasing such phase transition large pressure.

### 4- Mechanical properties from the elastic constants

The investigation of the mechanical properties was based on the calculation of the elastic properties determined by performing finite distortions of the lattice and deriving the elastic constants from the strain-stress relationship. The calculated sets of elastic constants $C_{ij}$ (i and j indicate directions) are given in Table 2. All $C_{ij}$ values are positive signaling stability of the two allotropes. The products obey the stability rules regarding the orthorhombic system:

$C_{ii}$ (i =1, 4, 5, 6) > 0;

$C_{11}C_{22} - C_{12}^2 > 0$;

$C_{11}C_{22}C_{33} + 2C_{12}C_{13}C_{23} - C_{11}C_{23}^2 - C_{22}C_{13}^2 - C_{33}C_{12}^2 > 0$.

Using ELATE program introduced above [18], the bulk and the shear modules obtained by averaging the elastic constants using Voigt's [19] method are also given, namely $B_V$ and $G_V$.

The bulk modules observed for **tfi** and $4^4T39$ $C_{12}$ are found in reasonable agreement with the EOS fit values in Fig. 4. The largest magnitude is observed for the latter with $B_V$= 428 GPa, but remains lower than diamond $B_V$ =444 GPa; this also applies for the shear modulus which amounts to $G_V$= 499 GPa versus $G_V$= 534 GPa for diamond [26]. A determining criterion for hardness is the corresponding Pugh ratio $G_V/B_V$ [27], relevant to ductile to brittle criteria. For $G_V/B_V < 1$ a trend to ductile behavior is expected whereas for $G_V/B_V > 1$ a brittle behavior is deducted. Indeed, **tfi** $C_{12}$ is identified as ductile due to $G_V/B_V = 0.60$ wherein C=C play a major role. Oppositely, $4^4T39$ $C_{12}$ has a larger value 1.17 pertaining to large brittleness. Such trends are translated into differences of Vickers hardness Hv calculated along with two models of microscopic theory of hardness:

$H_V^1= 0.92(G_V/B_V)^{1.137} G_V^{0.708}$ (Tian et al.) [20]

$H_V^2= 2(G_V^3/B_V^2)^{0.585} - 3$ (Chen et al.) [21]

The corresponding Vickers hardness ($H_V$) magnitudes obtained with the two methods are given in the last two columns of Table 2. They show close magnitudes letting confirm soft **tfi** $C_{12}$ versus ultra-hard $4^4\textbf{T}39$ $C_{12}$. Nevertheless, for the latter the obtained Vickers hardness magnitude remains smaller than diamond's which amounts to $H_V$=96 GPa [26].

5- **Dynamic properties**

To verify the dynamic stability of the carbon allotropes, an analysis of their phonon properties was performed. The phonon band structures obtained from a high resolution of the orthorhombic Brillouin zone BZ in accordance with the method proposed by Togo *et al*. [22] are shown in Figure 5. The bands (red lines) develop along the main directions of the orthorhombic Brillouin zone (horizontal *x*-axis), separated by vertical lines for better visualization, while the vertical direction (*y*-axis) represents the frequencies ω, given in terahertz (THz). For each crystal system the phonons band structures include 3N bands (N number of atoms) describing three acoustic modes starting from zero energy (ω = 0) at the Γ point (the center of the Brillouin zone) and reaching up to a few terahertz, and 3N-3 optical modes at higher energies. The low-frequency acoustic modes are associated with the rigid translation modes (two transverse and one longitudinal) of the crystal lattice. The calculated phonon frequencies are all positive, indicating that the allotropes are dynamically stable. The highest bands are observed for **tfi** $C_{12}$ around ~45-50 THz due to the C=C 1640 - 1680 $cm^{-1}$. Since 1 $cm^{-1}$ = 0.03 THz, one obtains 49.2 - 50.4 THz in agreement with the range of Fig. 5a highest frequencies. Concomitantly with the loss of $sp^2$ character the phonons panel of the other allotrope does not show high frequencies. The other bands remain below 40 THz the value observed for diamond by Raman spectroscopy [28]. This can be assessed by the fact that the distorted tetrahedral $sp^3$ network leaves less relationship with diamond for both allotropes.

6- **Electronic band structures**

Using the ground state crystal structure parameters in Table 1, the electronic band structures were obtained for the two carbon allotropes using the all-electrons DFT-based augmented spherical wave method (ASW) [23] and GGA XC approximation [13]. The band structures are displayed in Figure 6. The bands develop along the main directions of the primitive orthorhombic Brillouin zone. Along the vertical direction the energy zero is with respect to the Fermi level in Fig. 6a signaling a metallic behavior for **tfi** $C_{12}$. Oppositely Fig. 6b shows an energy gap signaling a semi-conducting behavior with a small energy gap < 1eV.

The zero energy is then considered at $E_V$ defining the top of the valence band (VB) separated from the higher energy conduction band (CB). Then the phase transition induces the passage from metallic to semi-conducting behavior for orthorhombic $C_{12}$.

**Conclusion**

Based on quantum mechanics calculations of ground state crystal structures and pertaining physical properties, we presented two novel orthorhombic carbon allotropes with $C_{12}$ stoichiometry. Specifically, the structures were found with distorted *C4* tetrahedra connected with C-C segments that change their behavior form $sp^2$ in NP **tfi** $C_{12}$ to the loss of the trigonal character in HP $C_{12}$ characterized with $C(sp^3)$ only and $4^4\mathbf{T}39$ topology. The quadratic fit of the E(V) energy-volume curves with Birch equations of states (EOS) to the $3^{rd}$ order let define the equilibrium $NP(E_0,V_0)$ at lower energy and higher volume versus $HP(E_0, V_0)$ at higher energy and smaller volume with an estimated transition pressure of ~100 GPa. Both allotropes found cohesive and mechanically stable were found with low and large Vickers hardness magnitudes: $H_V(\textbf{tfi } C_{12})$ =24 GPa and $H_V(4^4\mathbf{T}39\ C_{12})$ =90 GPa. From phonons band structure characterized by positive acoustic and optic frequencies a signature of trigonal C=C frequencies in **tfi** $C_{12}$ was identified in agreement with experiment. A metallic behavior was identified for for **NP tfi** $C_{12}$, like the **tfi** archetype tetragonal $C_6$ "glitter" oppositely to HP $4,4\mathbf{T}39$ $C_{12}$ identified with a small band gap letting assign semiconducting properties. The work is meant to open further the scope of $C(sp^2) \rightarrow C(sp^3)$ transformation mechanisms.

**Tables**

Table 1. Crystal structural properties of two allotropes $C_{12}$ in base centered orthorhombic space groups.

| Allotrope topology | **tfi $C_{12}$** $Ama2$ No. 40 | $4^4T39$ $C_{12}$ $Aea2$ No. 41 |
|---|---|---|
| a, Å | 5.9801 | 5.6448 |
| b, Å | 3.7300 | 4.5358 |
| c, Å | 3.5714 | 2.5710 |
| Shortest dist. Å | 1.53 | 1.59 |
| Angle ∠C-C-C ° | 105 | 107 |
| Volume, Å$^3$ | 79.66 | 65.83 |
| V/at., Å$^3$ | 6.64 | 5.49 |
| Density ρ (g/cm$^3$) | 3.00 | 3.63 |
| Atomic positions | C1 (4$a$) ½, ½, 0.90  C2 (8$c$) 0.1381, ¼, 0.1500 | C1 (4$a$) ½, ½, 0.90  C2 (8$b$) 0.3329, ¼, 0.1500 |
| $E_{total}$, eV | -103.41 | -93.65 |
| $E_{tot.}$/at., eV | -8.62 | -7.80 |
| $E_{coh}$/at., eV | -2.02 | -1.20 |

Table 2. Orthorhombic $C_{12}$ allotropes: Elastic Constants and Voigt-average properties of bulk $B_V$ and shear $G_V$ modules. All values are in GPa units.

| $C_{ij}$ | $C_{11}$ | $C_{22}$ | $C_{12}$ | $C_{13}$ | $C_{33}$ | $C_{44}$ | $C_{55}$ | $C_{66}$ | $B_V$ | $G_V$ | $G_V/B_V$ | $H_V^1$ | $H_V^2$ |
|---|---|---|---|---|---|---|---|---|---|---|---|---|---|
| **tfi** | 1171 | 462 | 98 | 98 | 461 | 88 | 337 | 88 | 345 | 208 | 0.60 | 23 | 24 |
| $4^4T39$ | 1122 | 1096 | 95 | 30 | 1206 | 495 | 483 | 445 | 428 | 499 | 1.17 | 89 | 90 |

FIGURES

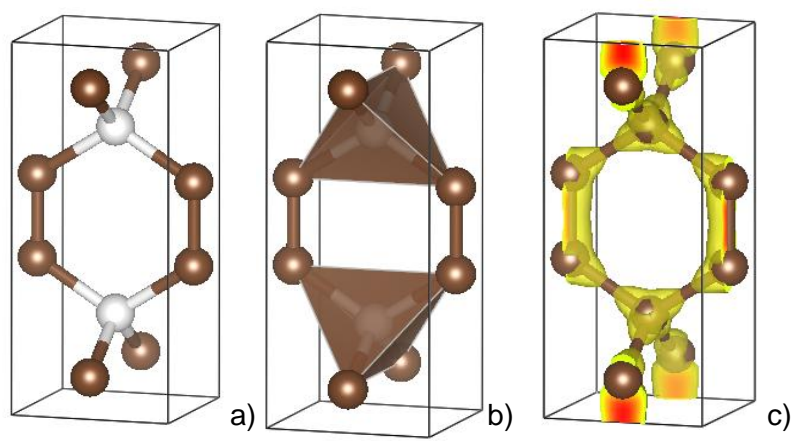

Figure 1. Projections along c tetragonal axis of $C_6$ glitter [6]: a) ball-and-stick, b) polyhedral, and c) charge density volumes (yellow) and planes (red) (cf. text)

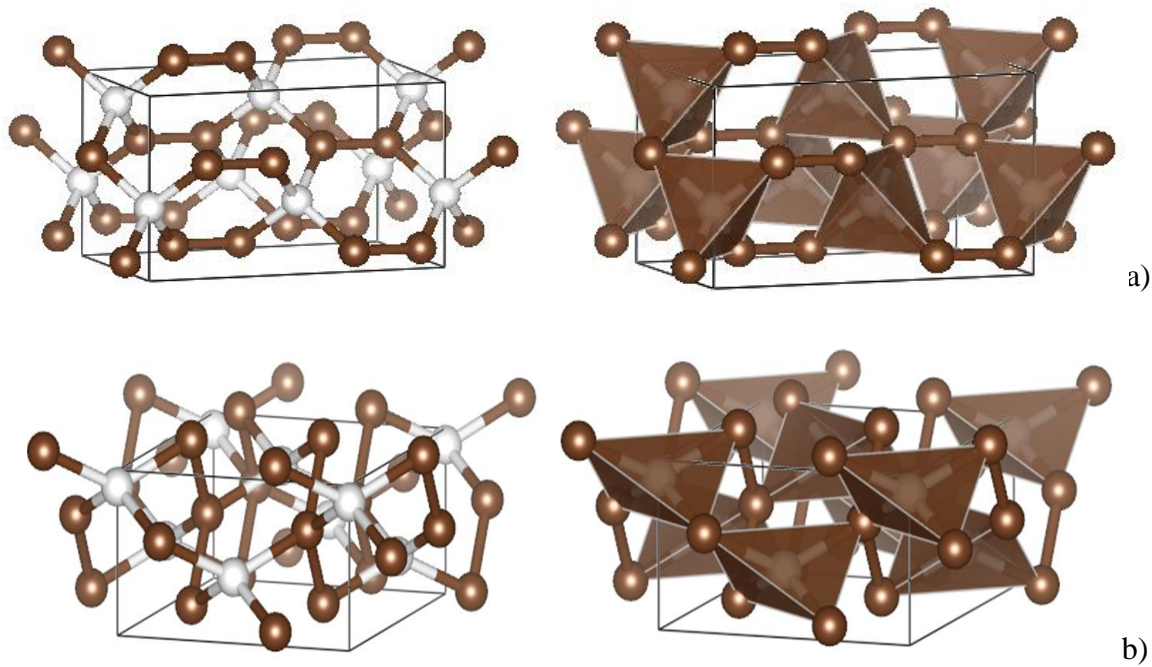

Figure 2. Ball-and-stick and polyhedral projections along c axis of orthorhombic **tfi** C$_{12}$ glitter (a), and 4$^4$**T**39 C$_{12}$ (b).

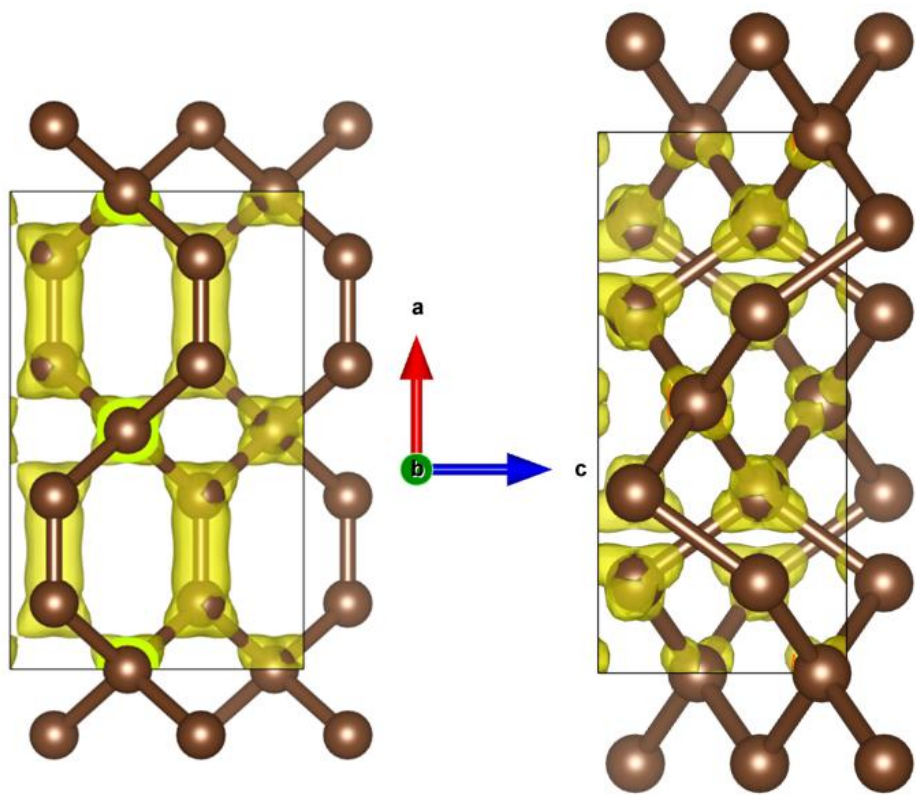

Figure 3. Charge density volumes (yellow) of **tfi** $C_{12}$ exhibiting the C=C-like segments beside tetrahedral volumes (left), and $4^4\mathbf{T}39$ $C_{12}$ showing solely tetrahedral-like charge densities (right). Note the passage from parallel C-C (left) to cross-like C-C (right).

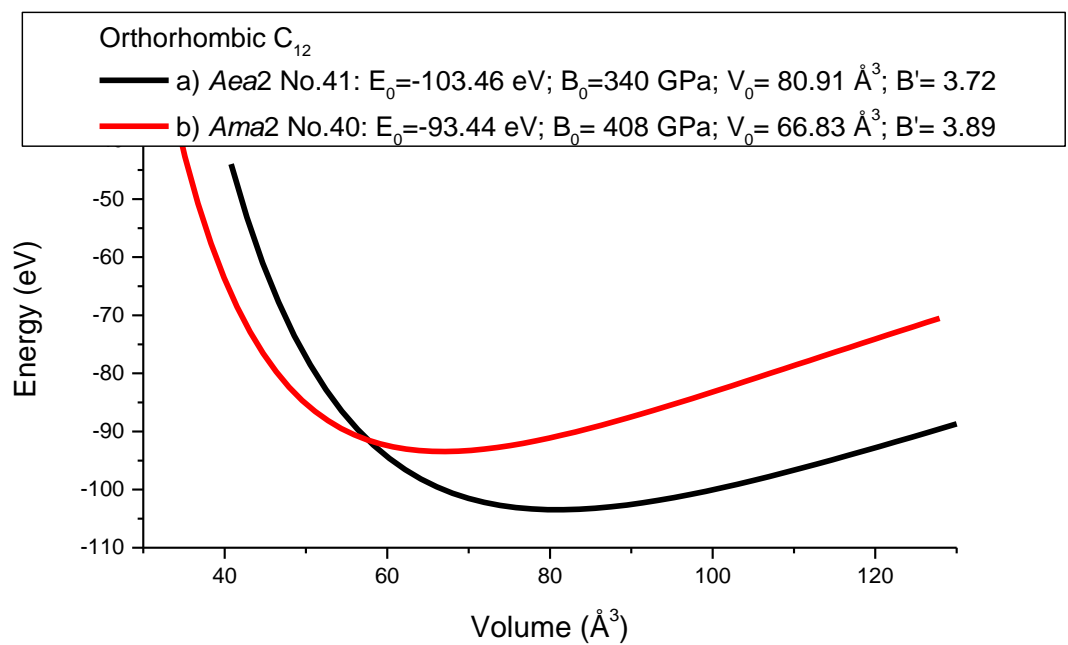

Figure 4. Energy -volume curves and 3$^{rd}$ order Birch equation of state (EOS) fit values of a) **tfi** $C_{12}$ and b) $4^4$**T**39 $C_{12}$.

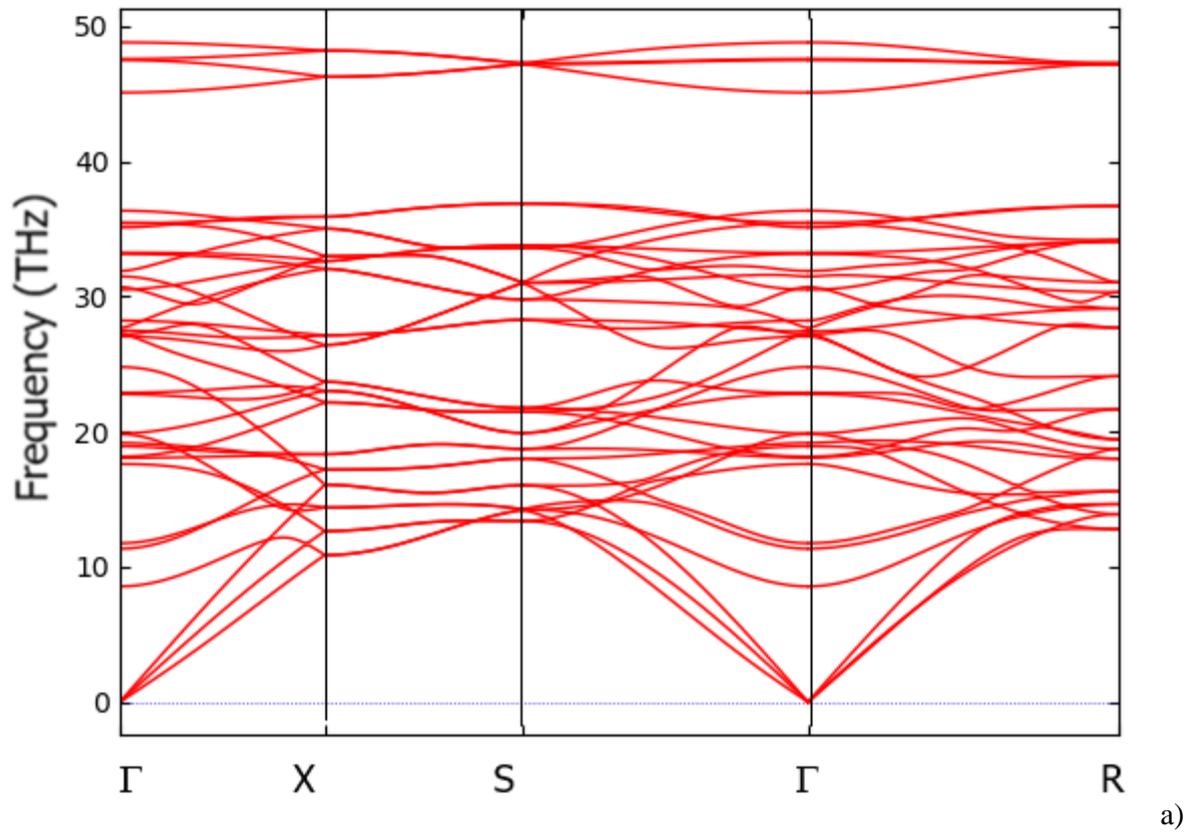

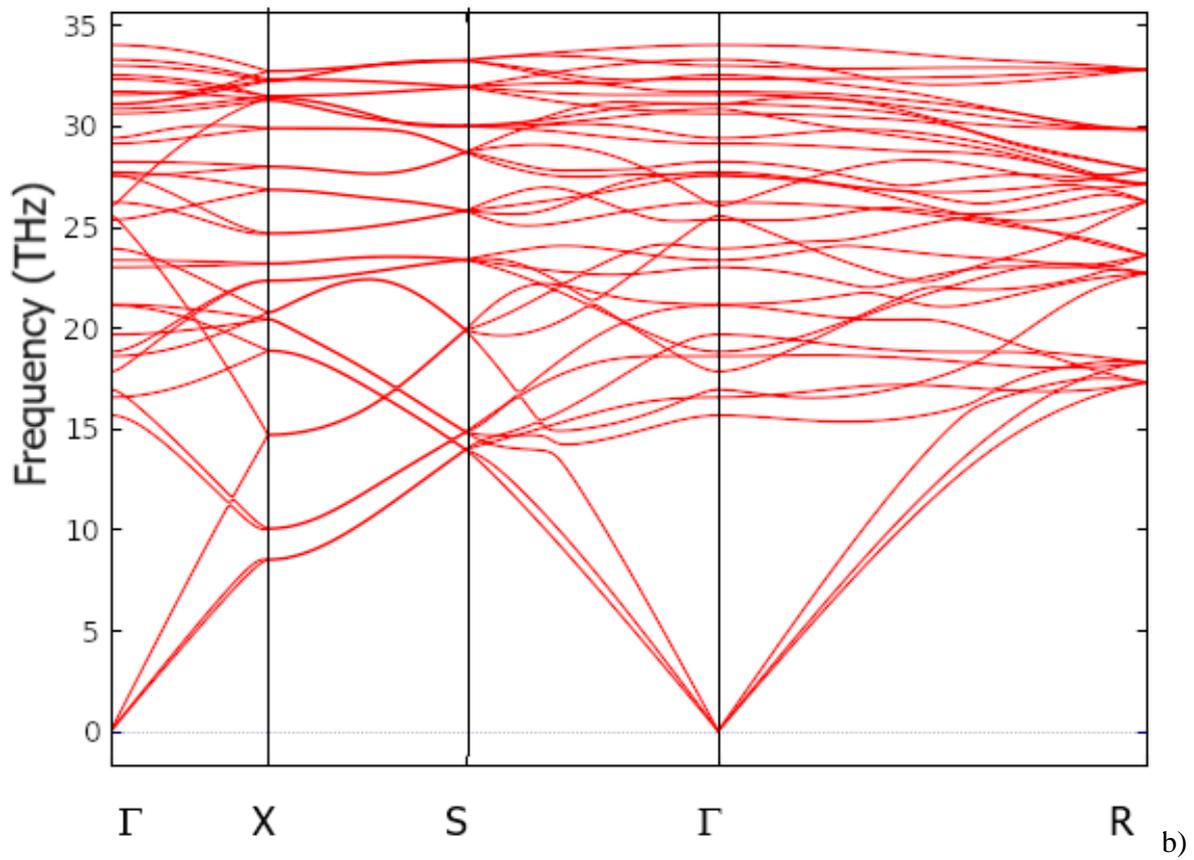

Figure 5. Phonon band structures along the major lines of the orthorhombic Brillouin zone. a) a) **tfi** $C_{12}$ and b) $4^4$**T**39 $C_{12}$

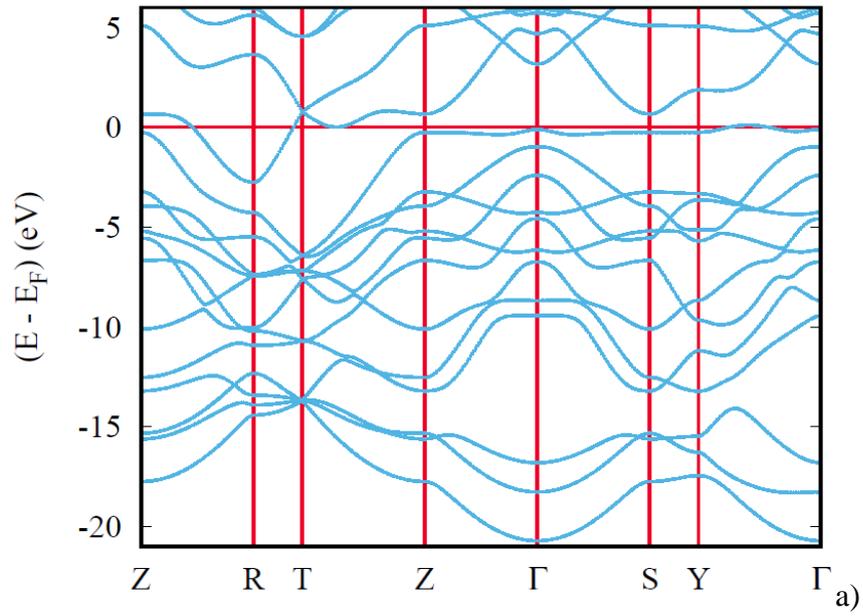

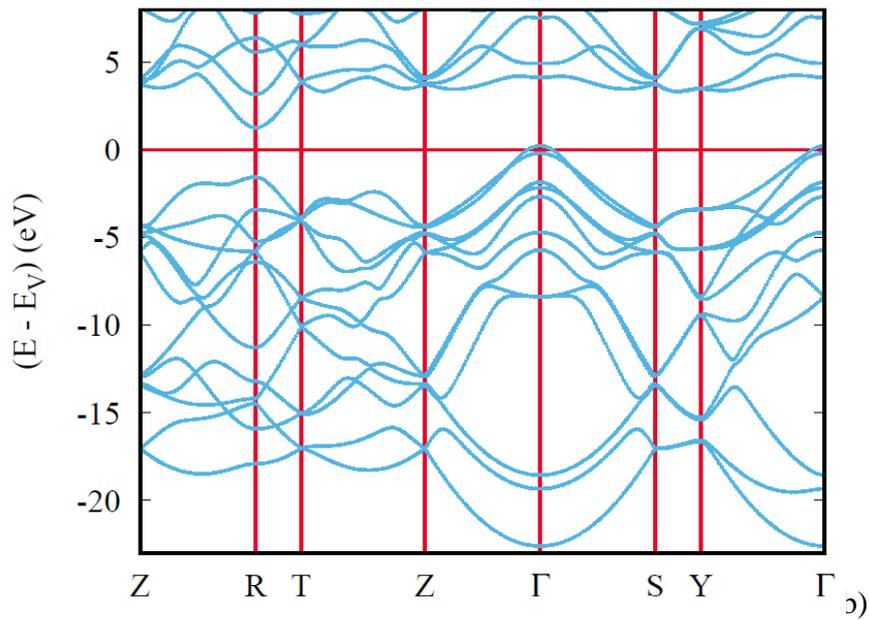

Figure 6. Electronic band structures. a) **tfi** $C_{12}$, and b) $4^4\mathbf{T}39$ $C_{12}$.